\documentstyle[preprint,aps,epsfig]{revtex}

\begin{document}

\title{
\rightline{\small MPI-PhT/2000-09}
Either neutralino dark matter or cuspy dark halos}

\author{Paolo Gondolo\thanks{Electronic address: {\tt gondolo@mppmu.mpg.de}}}

\address{Max-Planck-Institut f\"{u}r Physik, F\"{o}hringer Ring 6, D-80805
  M\"{u}nchen, Germany}

\maketitle

\begin{abstract}
  We show that if the neutralino in the minimal supersymmetric standard model
  is the dark matter in our galaxy, there cannot be a dark matter cusp
  extending to the galactic center. Conversely, if a dark matter cusp extends
  to the galactic center, the neutralino cannot be the dark matter in our
  galaxy.  We obtain these results considering the synchrotron emission from
  neutralino annihilations around the black hole at the galactic center.
\end{abstract}

The composition of dark matter is one of the major issues in cosmology. A
popular candidate for non-baryonic cold dark matter is the lightest neutralino
appearing in a large class of supersymmetric models~\cite{jun96}.
In a wide range of supersymmetric parameter space, relic neutralinos from the
Big Bang are in principle abundant enough to account for the dark matter in our
galactic halo~\cite{eds97}. 

A generic prediction of cold dark matter models is that dark matter halos
should be have steep central cusps, meaning that their density rises as
$r^{-\gamma}$ to the center.  Semi-analytical calculations find a cusp slope
$\gamma$ between $\sim 1$~\cite{sw98} and 2~\cite{hs85}. Simulations find a
slope $\gamma$ ranging from 0.3~\cite{kra98} to 1~\cite{nfw} to
1.5~\cite{fuk97}.  It is unclear if dark matter profiles in real galaxies and
galaxy clusters have a central cusp or a constant density core.

There is mounting evidence that the non-thermal radio source Sgr A$^*$ at the
galactic center is a black hole of mass $ M \sim 3 \times 10^6 \, M_{\odot}$.
This inference is based on the large proper motion of nearby
stars~\cite{ghe98}, the spectrum of Sgr A$^*$ (e.g.~\cite{mez96,nar98}), and
its low proper motion~\cite{bac99}. It is difficult to explain these data
without a black hole~\cite{mao98}. 

The black hole at the galactic center modifies the distribution of dark matter
in its surroundings~\cite{gs99}, creating a high density dark matter region
called the spike -- to distinguish it from the above mentioned cusp. Signals
from particle dark matter annihilation in the spike may be used to discriminate
between a central cusp and a central core.  With a central cusp, the
annihilation signals from the galactic center increase by many orders of
magnitude. With a central core, the annihilation signals do not increase
significantly.

Stellar winds are observed to pervade the inner parsec of the
galaxy~\cite{mez96}, and are supposed to feed the central black hole
(e.g.~\cite{nar98,cok99}). These winds carry a magnetic field whose measured
intensity is a few milligauss at a distance of $\sim 5{\rm pc}$ from the
galactic center~\cite{yus96}. The magnetic field intensity can rise to a few
kilogauss at the Schwarzschild radius of the black hole in some accretion
models for Sgr A$^*$~\cite{mel92}.

In this letter we examine the radio emission from neutralino dark matter
annihilation in the central spike. (Previous studies of radio emission from
neutralino annihilation at the galactic center have considered an $r^{-1.8}$
cusp but no spike~\cite{ber94}.)  Radio emission is due to synchrotron
radiation from annihilation electrons and positrons in the magnetic field
around Sgr A$^*$. Comparing the radio emission from the neutralino spike with
the measured Sgr A$^*$ spectrum, we find that neutralino dark matter in the
minimal supersymmetric standard model is incompatible with a dark matter cusp
extending to the galactic center.

There are two ways to interpret our results. If we believe that there is a dark
matter cusp extending to the center of our galaxy, we can exclude the
neutralino as a dark matter candidate. Conversely, if we believe that dark
matter is the lightest neutralino, we can exclude that a dark matter cusp
extends to the center of the galaxy.

{\it Dark matter candidate.} We examine the lightest neutralino in the minimal
supersymmetric standard model. This model provides a well-defined calculational
framework, but contains at least 106 yet-unmeasured parameters~\cite{dim95}.
Most of them control details of the squark and slepton sectors, and are usually
disregarded in neutralino dark matter studies (cfr.~\cite{jun96}). So,
following Bergstr\"om and Gondolo~\cite{ber96}, we restrict the number of
parameters to 7.  Out of the database of points in parameter space built in
refs.~\cite{eds97,ber96,ber98}, we use the 35121 points in which the neutralino
is a good cold dark matter candidate~\cite{eds97}, in the sense that its relic
density satisfies $0.025 < \Omega_\chi h^2 < 1 $.  The upper limit comes from
the age of the Universe, the lower one from requiring that neutralinos are a
major fraction of galactic dark halos.  Present understanding of the matter
density in the universe (e.g.~\cite{bah99}) suggests a narrower range $ 0.08 <
\Omega_\chi h^2 < 0.18 $, but we conservatively use the broader range.

{\it Spike profile.} We summarize the results of ref.~\cite{gs99} for the
spike profile. We assume the cusp has density profile
\begin{equation}
\label{eq:cusp}
\rho_{\rm cusp} = \rho_D \left( \frac{r}{D} \right)^{-\gamma} ,
\end{equation}
with $\rho_D=0.24{\rm GeV}/c^2/{\rm cm}^3$ the density at the reference point
$D=8.5{\rm kpc}$, the Sun location (this is a conservative value for $\rho_D$,
see~\cite{gs99}). Then within a central region of radius
$
R_{\rm sp} = \alpha_{\gamma} D \left( M/\rho_D D^3
\right)^{1/(3-\gamma)} ,
$
where $\alpha_{\gamma}$ is given in ref.~\cite{gs99} and
$M=(2.6 \pm 0.2) \times 10^6 \, M_{\odot}$ is the mass of the central black
hole, 
the dark matter density is modified to
\begin{equation}
\rho_{\rm sp} = \frac{ \rho'(r) \rho_{\rm c} } { \rho'(r) + \rho_{\rm
    c} } .
\end{equation}
Here $ \rho_{\rm c} = m_\chi/(\sigma v t_{\rm bh}) $, where $t_{\rm bh}$ is the
age of the black hole (conservatively $10^{10}$ yr), $m_\chi$ is the mass of
the neutralino, and $\sigma v$ is the neutralino--neutralino annihilation cross
section times relative velocity (notice that for neutralinos at the galactic
center $\sigma v$ is independent of $v$).  Furthermore,
\begin{equation}
\rho'(r) = 
\rho_R \, g(r) \, 
\left( \frac{R_{\rm sp}}{r}\right)^{\gamma_{\rm sp}} ,
\end{equation}
with $ g(r) = \left[ 1 - (8GM)/(rc^2) \right]^3 $ accounting for dark matter
capture into the black hole, $ \gamma_{\rm sp} = (9-2\gamma)/(4-\gamma) $,
and $\rho_R=\rho_D \left( {R_{\rm sp}}/{D} \right)^{-\gamma}$.

{\it Annihilation rate.} The total number of neutralino annihilations per
second in the spike follows from the density profile as
\begin{equation}
\Gamma = \frac{\sigma v}{m^2} \int \rho_{\rm sp}^2 4 \pi r^2 dr =
\frac{ 4\pi \sigma v \rho_{\rm in}^2 R_{\rm in}^3} {m^2} ,
\end{equation}
with $\rho_{\rm in} = \rho_{\rm sp}(R_{\rm in})$ and $R_{\rm in} = 1.5 \left[
  (20 R_{\rm S})^2+R_{\rm c}^2 \right]^{1/2} $. The latter expression is a good
approximation (6\%) to the numerical integration of the annihilation profile.

Most of the annihilations occur either close to the black hole at $\sim 13
R_{\rm S} \sim 3 \times 10^{-6} {\rm pc}$ (where $R_{\rm S}\equiv 2GM/c^2$ is
the Schwarzschild radius) or around the spike core radius $R_{\rm c} = R_{\rm
  sp} \left( \rho_R/\rho_{\rm c} \right) ^ {1/\gamma_{\rm sp}} $, whichever is
larger.

{\it Radio signals.}  The electrons and positrons produced by neutralino
annihilation in the spike are expected to emit synchrotron radiation in the
magnetic field around the galactic center.

The strength and structure of this magnetic field is known to some extent.  A
magnetic field of few milligauss has been detected~\cite{yus96} few parsecs
from the center. Models of Sgr A$^*$ contain accretion flows, either
spherical~\cite{mel92} or moderately flattened~\cite{nar98}, which carry a
magnetic field towards the black hole.  The strength of this magnetic field is
assumed to increase inwards according to magnetic flux conservation or
equipartition.

Including the gas and the radial dependence of the magnetic field in the
synchrotron emission from neutralino annihilations is a complicated problem.
Electrons and positrons in the regions where the magnetic field is strong may
lose their energy almost in place, while those at the outskirts of the spike
may have time to diffuse to very different radii.  Moreover, the plasma may
affect the shape of the synchrotron spectrum. We postpone this complicated
analysis, and consider three simple but relevant models for the magnetic field
and the electron/positron propagation.

In model A, we assume that the magnetic field is uniform across the spike,
with strength $B=1{\rm mG}$, and that the electrons and positrons lose all
their energy into synchrotron radiation without moving significantly from their
production point.

In model B, we also assume that the magnetic field is uniform across the spike
with strength $B=1{\rm mG}$, but that the electrons and positrons diffuse
efficiently and are redistributed according to a gaussian encompassing the
spike (we take the gaussian width $\lambda=1$ pc).

In model C, we assume that the magnetic field follows the equipartition value
$B = 1 \mu{\rm G} (r/{\rm pc})^{-5/4}$ (from ref.~\cite{mel92}) and that the
electrons and positrons lose all their energy into synchrotron radiation
without moving significantly from their production point. In addition, in this
model, we
neglect synchrotron self-absorption.

Under these assumptions, the electron plus positron spectrum follows from the
equation of energy loss $-dE/dt = P(E) \equiv (2e^4B^2E^2)/(3m_e^4c^7)$ as
\begin{equation}
  \frac{d n_e}{dE} = \frac{Y_e(\mathord{>}E)}{P(E)} \, \Gamma \, f_e(r) ,
\end{equation}
where 
\begin{equation}
  f_e(r) = \frac{\rho_{\rm sp}^2}{\int \rho_{\rm sp}^2 4 \pi r^2 dr}
\end{equation}
in models A and C, and
\begin{equation}
  f_e(r) = \frac{1}{(2\pi\lambda^2)^{3/2}} e^{-r^2/2\lambda^2}
\end{equation}
in model B.

$Y_e(\mathord{>}E)$ is the number of annihilation electrons and positrons with
energy above $E$. We obtain $Y_e(\mathord{>}E)$ with the DarkSUSY
code~\cite{DarkSUSY}, which includes a Pythia simulation of the $e^\pm$
continuum and the $e^\pm$ lines at the neutralino mass~\cite{bal99}.

The synchrotron luminosity is given by
\begin{equation}
\label{eq:Snu}
  L_{\nu} = \frac{ A_{\nu} \Gamma}{\nu} \,
  \int dr 4\pi r^2 f_e(r) \, \int_{m_e}^m \frac{Y_e(\mathord{>}E)}{\nu_c(E)} \,
  F\!\left(\frac{\nu}{\nu_c(E)}\right) dE ,
\end{equation}
where
\begin{equation}
  \nu_c(E) = \frac{3eB}{4\pi m_e c} \left(\frac{E}{m_e c^2} \right)^2
\end{equation}
and
\begin{equation}
  F(x) = \frac{9\sqrt{3}}{8\pi} x \int_x^\infty K_{5/3}(y) dy .
\end{equation}
The factor $A_{\nu}$ accounts for synchrotron self-absorption. In models A and
B, we write
\begin{equation}
A_{\nu} = \frac{1}{a_{\nu}} \int_0^\infty \left[ 1 - e^{-\tau(b)}
\right] \pi b db ,
\end{equation}
where $(b,z)$ are cylindrical coordinates,
\begin{equation}
\tau = a_{\nu} \int_{-\infty}^{+\infty} f_e(b,z) dz ,
\end{equation}
and
\begin{equation}
a_{\nu} = \frac{e^3 B \Gamma}{9 m_e \nu^2} \int_{m_e}^m E^2
\frac{d}{dE} \! \left( \frac{Y_e(\mathord{>}E)}{E^2 P(E)}\right) \, 
F\!\left(\frac{\nu}{\nu_c(E)} \right) dE .
\end{equation}
In model C, we neglect self-absorption ($A_\nu=1$).

We have evaluated equation~(\ref{eq:Snu}) numerically for each point in
supersymmetric parameter space. In model C, we use the approximation $F(x)
\simeq \delta(x-0.29)$, which selects the peak of the synchrotron emission from
each electron or positron (profuse thanks to Pasquale Blasi for suggesting this
approximation).

Figure 1 shows a comparison of typical synchrotron spectra from neutralino
annihilation in the spike with the measured spectrum of Sgr A$^*$ (the latter
is taken from the compilation in ref.~\cite{nar98}). Four spectra are plotted,
corresponding to two points in supersymmetric parameter space (thick and thin
lines) and two assumptions for the magnetic field (solid and dashed lines; for
models A and C, respectively). The spectra are normalized to their maximal
intensity, which is fixed by the upper bound at 408 MHz~\cite{dav76}. This
upper bound limits the synchrotron intensity for all points in supersymmetric
parameter space.

{\it Results.}  If a dark matter cusp extends to the galactic center, the
neutralino cannot be the dark matter in our galaxy. For example, let us assume
that the halo profile is of the Navarro-Frenk-White form~\cite{nfw}, namely
$\rho \propto r^{-1}$ in the central region. Figure 2 shows the expected radio
fluxes $S_\nu = L_\nu/4\pi D^2$ at 408 MHz and the upper limit
from~\cite{dav76}. The upper panel is for model A, the lower panel for model C.
Results of model B are similar to those of model A. Irrespective of the
assumption on the magnetic field or the $e^\pm$ propagation, all points in
supersymmetric parameter space where the neutralino would be a good dark matter
candidate are excluded by several orders of magnitude.

Conversely, if the neutralino is the dark matter, there is no steep dark matter
cusp extending to the galactic center. We see this by lowering the cusp slope
$\gamma$ until the expected flux at 408 MHz decreases below the upper limit.
We obtain a different maximum value $\gamma_{\rm max}$ for each point in
supersymmetric parameter space. These values are plotted in figure 3 together
with the range $0.3 \lesssim \gamma \lesssim 1.5$ obtained in cold dark matter
simulations. The upper bounds $\gamma_{\rm max}$ are generally orders of
magnitude smaller than the simulation results.

We conclude that neutralino dark matter in the minimal supersymmetric standard
model is incompatible with a dark matter cusp extending to the galactic center.
If there is a dark matter cusp extending to the center, we can exclude the
neutralino in the minimal supersymmetric standard model as a dark matter
candidate.  Conversely, if the dark matter of the galactic halo is the lightest
neutralino in the minimal supersymmetric standard model, we can exclude that a
dark matter cusp extends to the center of the galaxy.

{\it Acknowledgements.} Many thanks to the Fermilab Astrophysics group for the
generous and warm hospitality. Thanks in particular to Pasquale Blasi for
insistingly requesting a non-uniform magnetic field (model C).

\begin{figure}
\label{fig1}
\begin{center}
\epsfig{width=\textwidth,file=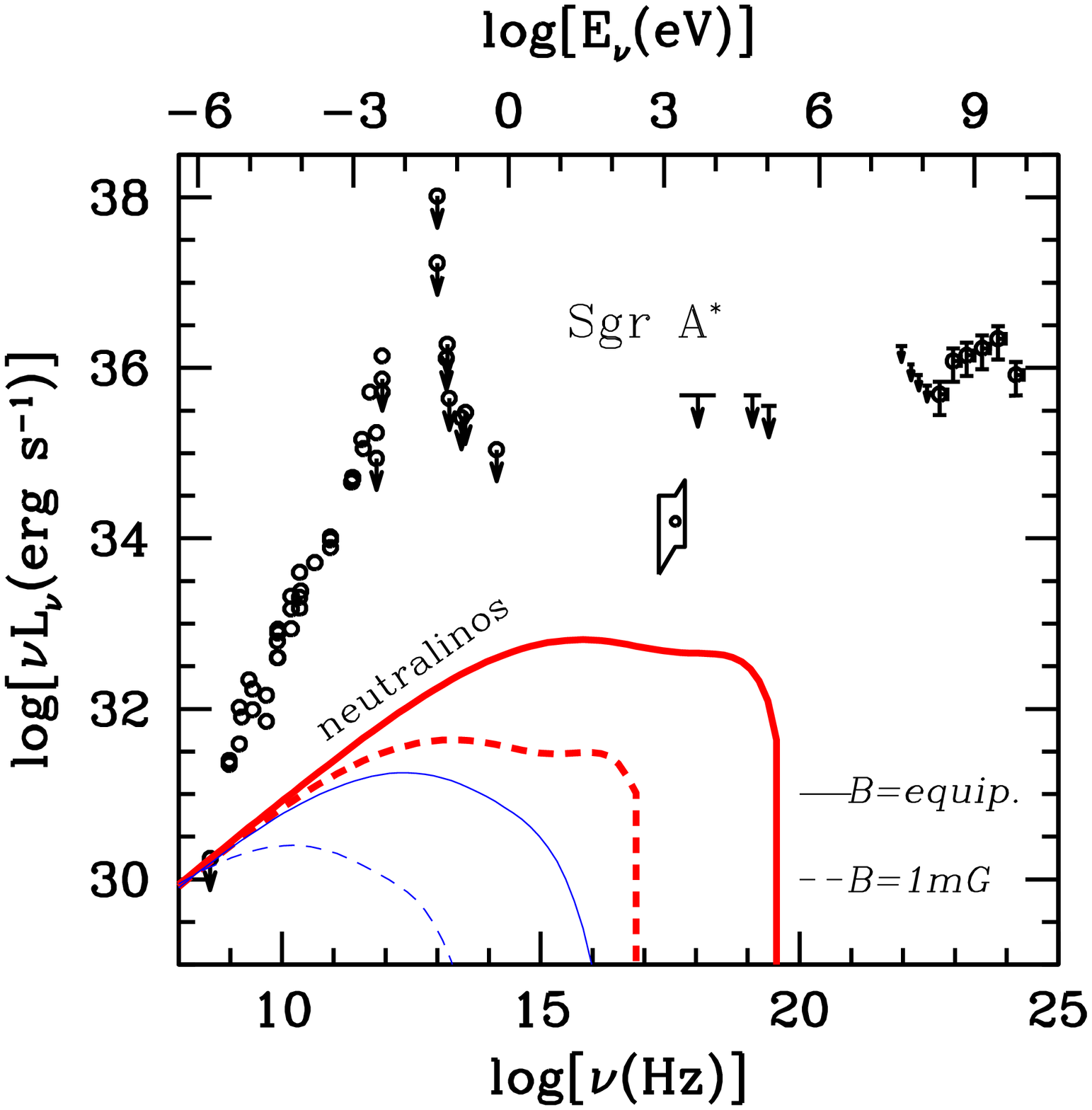}
\end{center}
\caption{
  Comparison of the Sgr A$^*$ spectrum with the synchrotron emission from
  neutralino annihilation in the spike. The figure shows four typical
  synchrotron spectra: two points in supersymmetric parameter space (thick and
  thin lines), and two models for the magnetic field (solid and dashed lines).
  The spectra are normalized to the upper bound at 408 MHz. }
\end{figure}

\begin{figure}
\label{fig2}
\begin{center}
\epsfig{width=\textwidth,file=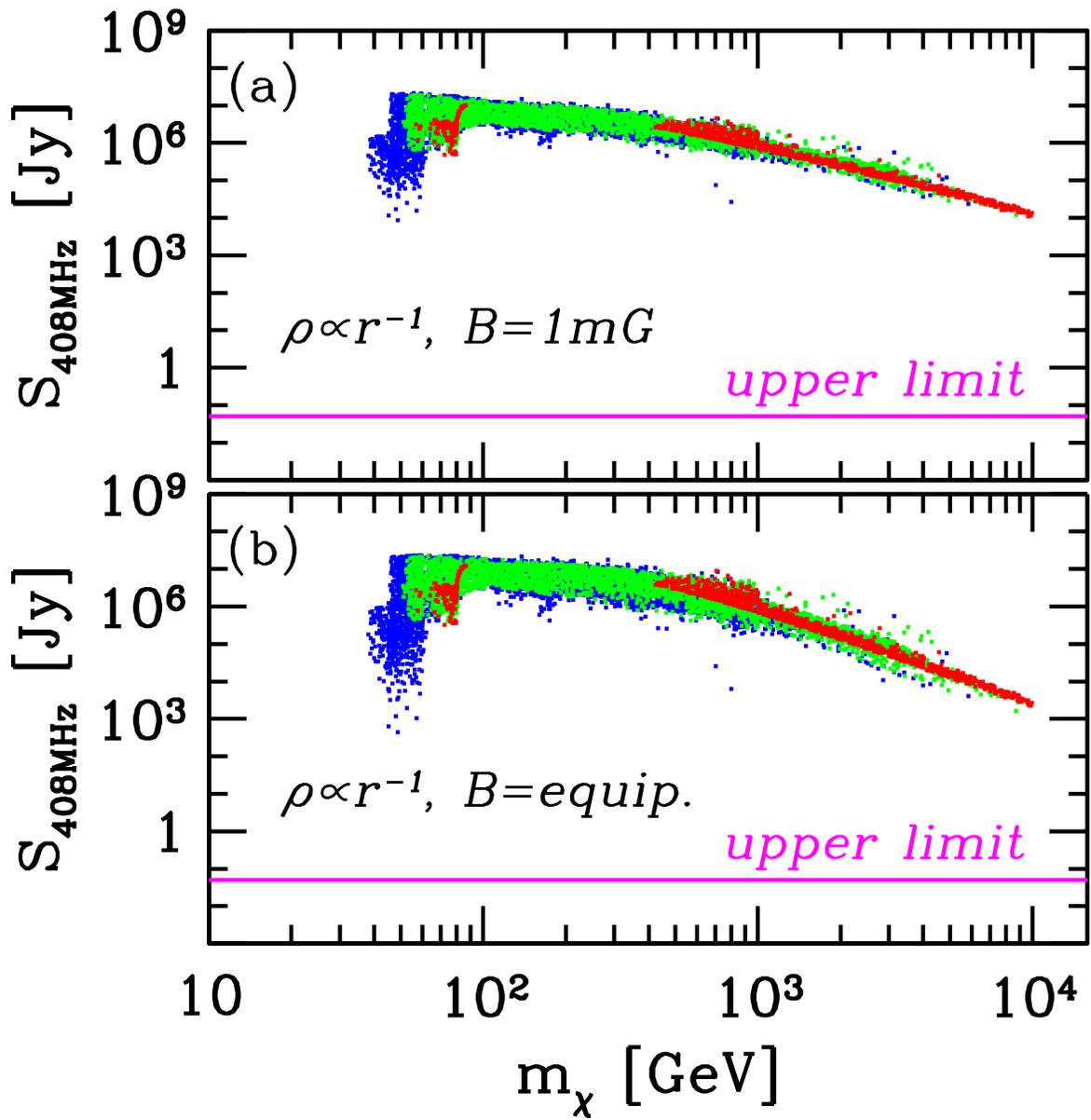}
\end{center}
\caption{
  Expected radio emission from the galactic center at 408 MHz from neutralino
  annihilations in the dark matter spike, assuming a Navarro-Frenk-White
  profile and (a) a uniform magnetic field of 1 mG, (b) a magnetic field at
  the equipartition value. All models exceed the present upper bound by several
  orders of magnitude.}
\end{figure}

\begin{figure}
\label{fig3}
\begin{center}
\epsfig{width=\textwidth,file=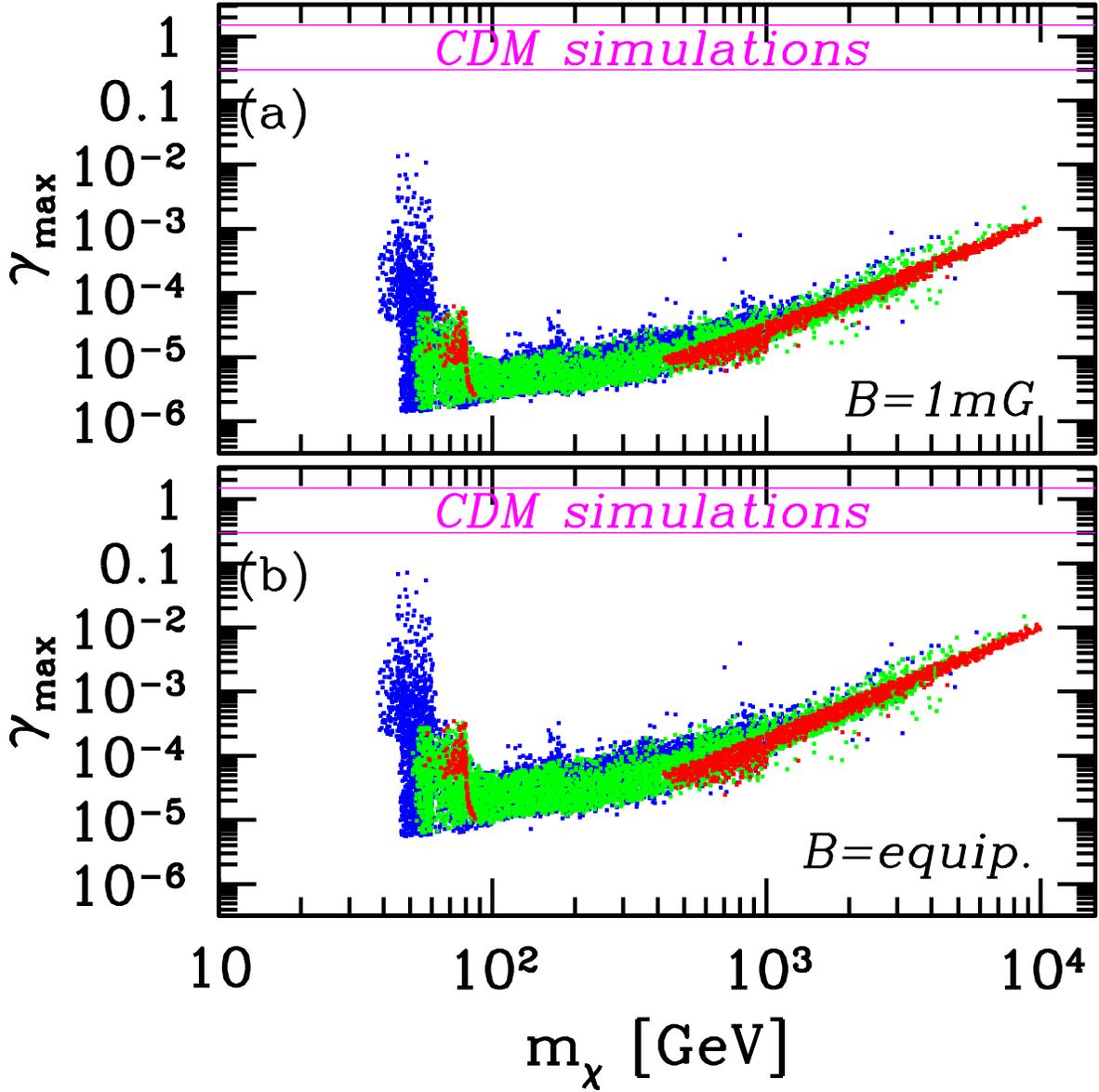}
\end{center}
\caption{
  Upper bound on the inner halo slope $\gamma$ imposed by the constraint on the
  radio emission from the galactic center at 408 MHz, assuming (a) a uniform
  magnetic field of 1 mG, and (b) a magnetic field at the equipartition
  value. Each dot corresponds to a point in supersymmetric parameter space.
  The results of cold dark matter simulations are much higher than the
  upper bounds.}
\end{figure}

\end{document}